\documentclass[floatfix,aps,prb,twocolumn,superscriptaddress,
10pt]{revtex4-2}

\usepackage{graphicx}
\usepackage{ulem}
\usepackage{color}
\usepackage{physics}
\usepackage{booktabs}
\usepackage{array}
\usepackage{listings}
\usepackage{soul}
\usepackage{float}
\usepackage[T1]{fontenc}
\usepackage[utf8]{inputenc}
\usepackage{caption}
\usepackage{subcaption}
\usepackage{afterpage}
\usepackage{xcolor}
\usepackage{multirow}
\usepackage{diagbox}
\usepackage{hyperref}
\hypersetup{colorlinks=true, linkcolor=blue, citecolor=blue, urlcolor=blue}

\begin{document}

\widetext
\title{The anti-distortive polaron : an alternative mechanism for lattice-mediated charge trapping}

\author{Hamideh Hassani}

\affiliation{Physique Th\'eorique des Mat\'eriaux, QMAT, CESAM, Universit\'e de L\`iege, B-4000 Sart-Tilman, Belgium}
\affiliation{Department of Physics, University of Antwerp, Groenenborgerlaan 171, 2020 Antwerp, Belgium}

\author{Eric Bousquet}
\affiliation{Physique Th\'eorique des Mat\'eriaux, QMAT, CESAM, Universit\'e de L\`iege, B-4000 Sart-Tilman, Belgium}

\author{Xu He}
\affiliation{Physique Th\'eorique des Mat\'eriaux, QMAT, CESAM, Universit\'e de L\`iege, B-4000 Sart-Tilman, Belgium}

\author{Bart Partoens}
\affiliation{Department of Physics, University of Antwerp, Groenenborgerlaan 171, 2020 Antwerp, Belgium}

\author{Philippe Ghosez}
\affiliation{Physique Th\'eorique des Mat\'eriaux, QMAT, CESAM, Universit\'e de L\`iege, B-4000 Sart-Tilman, Belgium}

\date{\today}

\begin{abstract}

Polarons can naturally form in materials from the interaction of extra charge carriers with the atomic lattice. Ubiquitous, they are central to various topics and phenomena such as high-T$_c$ superconductivity, electrochromism, photovoltaics, photocatalysis or ion batteries. However, polaron formation remains poorly understood and mostly relies on few historical models such as Landau-Pekar, Fr\"olich, Holstein or Jahn-Teller polarons.  Here, from advanced first-principles calculations, we show that the formation of intriguing medium-size polarons in WO$_3$ does not fit with traditional models but instead arises from the undoing of distortive atomic motions inherent to the pristine phase, which lowers the bandgap through dynamical covalency effects.  We so introduce the innovative concept of {\it anti-distortive} polaron and rationalize it from a quantum-dot model. We demonstrate that anti-distortive polarons are generic to different families of compounds and clarify how this new concept opens concrete perspectives for a better control of the polaronic state and related properties.

\end{abstract}
\maketitle

\section{Introduction}
Polarons consist in the spatial localization of excess charge carriers in crystals, through their interaction with the surrounding atomic lattice and the subsequent deformation of the latter~\cite{Emin2012}. 
As such, polarons can be seen as quasiparticles consisting of an electron or a hole, dressed by a cloud of virtual phonons attached to and moving with it. 
Depending on the crystal and strength of the electron-phonon interaction, polarons can be large or small but will inevitably affect electronic properties, making the concept of polaron a central, timely, and ubiquitous topic in material science~\cite{Alexandrov2010}. As such, it is nowadays at the heart of active research in photovoltaics~\cite{Photovoltaics}, photocatalysis~\cite{Photocatalysis}, electrochromism~\cite{DEB2008245} or ion batteries~\cite{Batteries}. 

Anticipated by Landau~\cite{Landau:1933}, the concept of polaron was first coined by Pekar~\cite{Pekar1946}, from a model describing the interaction of an electron with a continuous polarizable medium. 
Following the same line of thought, Fr\"ohlich~\cite{frohlich1950xx,Frhlich1954} then provided a more rigorous quantum mechanical description, relying on long-range Coulombic electron-phonon interactions. 
The addition of an extra charge in a polar crystal lattice naturally induces dielectric screening from the activation of longitudinal optical polar phonons, yielding the formation of a polaron, the spatial extension of which depends on the dielectric constant of the medium. 
This typically applies to large polarons extending over multiple sites in ionic and polar crystals. 
Independently, Holstein~\cite{HOLSTEIN1959222,HOLSTEIN1959325} provided an alternative and more atomistic description, relying on short-range electron-phonon interactions involving acoustic phonons. 
Local elastic deformations can create a small quantum dot trapping the charge and can explain the formation of small polarons, usually confining the charge on one single site. 
Later, the concept of Jahn-Teller polaron was also introduced by Hock, Nickisch, and Thomas~\cite{Hock1983} and appeared as an alternative mechanism for small polaron formation in a system with Jahn-Teller active ions. 
In the presence of degenerated energy levels, the addition of an extra charge can activate a local Jahn-Teller distortion that splits the energy levels and stabilizes the charge on a given site. 
Such a concept of Jahn-Teller polaron is intimately linked to important phenomena such as colossal magnetoresistance in manganites~\cite{1997Natur.386..256T,PhysRevB.54.5405} or the high-T$_c$ superconductivity in cuprates~\cite{muller}. 
All these consecutive models are still the backbone of present polaron interpretations. 
They remain however more conceptual than real tools for quantitative predictions.

Polarons are nowadays also accessible to density functional theory (DFT) calculations, although their first-principles study remains challenging~\cite{Franchini2021}. 
A first limitation in direct calculations is the size of simulation boxes, restricting practical investigations to small polarons~\cite{pssb.201046289,PhysRevB.78.235104}. 
This drawback has been recently overcome with the development of an elegant formalism requiring only the computation of quantities within the primitive cell~\cite{PhysRevLett.122.246403}. 
Including all kinds of electron-phonon interactions, this approach provides moreover a unified formalism for addressing large and small polarons. 
A second difficulty is to properly reproduce charge localization using usual local exchange-correlation functionals, that include spurious self-interactions. 
However, more advanced self-interaction corrections~\cite{PhysRevB.92.075202,Falletta22-prl}, DFT+U~\cite{Falletta22-npj,oxyg-vac4-doped1}, and hybrid functionals~\cite{Eric-PhysRevResearch.2.012052,oxyg-vac4-doped1} often reproduce polarons in fair agreement with experimental data.
Thanks to this recent progress, there has been a renewed interest in polarons~\cite{Franchini2021}, with the hope that DFT could unravel meaningful new insights in polaron formation and important related phenomena.      

WO$_3$ is often considered as a paradigmatic polaronic compound~\cite{salje1997,Salje-Main-Bipolaron}. 
Although the unusual medium size and disk-like shape of its polarons were already revealed in the experimental work of Salje~\cite{Salje-Main-Bipolaron}, the origin of such polarons remains unclear. 
Here, relying on first-principles investigations, we show that classical polaron models do not strictly apply to WO$_3$. 
Instead, we introduce the concept of {\it anti-distortive} polaron, in which charge localization arises from the undoing of some inherent distortive motions. 
We rationalize our findings with a simple generic model and further demonstrate the generality of the concept and its concrete practical implications.

\noindent
\section{Results}

Structurally, WO$_3$ can be seen as an ABO$_3$ perovskite with missing A cation: its $Pm\bar{3}m$ cubic reference structure consists in a network of corner-shared oxygen octahedra, with W atoms at their centers~\cite{Howard2002}. 
On cooling, it exhibits a complex sequence of structural phase transitions toward a monoclinic ground state, originally considered as of $Pc$ symmetry ~\cite{salje1997,WOODWARD19979} but better assigned now as $P2_1/c$~\cite{Hamdi2016,Ati-PhysRevB.105.014107}. 
Electronically, it is a wide bandgap semiconductor~\cite{SIMCHI2014609}. 
In practice however, it is typically sub-stoichiometric (WO$_{3 - \delta}$), which makes it intrinsically $n$-doped.
Then, those extra electrons are known to form polarons and even bi-polarons, which are expected to be closely associated with the remarkable electrical, chromic and superconducting properties of WO$_3$~\cite{DEB2008245,NIKLASSON2004315,Salje20,Shengelaya20}.

Accurate simulation of a self-trapped single polaron in WO$_3$ from DFT has only been reported recently~\cite{Gerosa2016,Galli-18, Eric-PhysRevResearch.2.012052}. 
Those calculations properly captured the spatial extension of the polaron over a few unit cells, together with its unusual disk-like shape. 
Building on those seminal results, we start here with a careful analysis of the polaron structural distortion in order to unravel the microscopic mechanism behind its formation.    

Our investigations are performed in the framework of DFT using a full hybrid-functional approach on large supercells (up to 576 atoms), which was previously shown to provide an accurate description of polarons in WO$_3$~\cite{Eric-PhysRevResearch.2.012052}. 
To study polaron formation, an extra electron is added to the system while charge neutrality is restored by the inclusion of a positive background (see Appendix.~A). The atomic distortion is then analyzed from its projection on phonon eigendisplacements of the reference ($Pm\bar{3}m$) and ground-state ($P2_1/c$) phases.

The reference $Pm\bar{3}m$ cubic phase of WO$_3$ shows various phonon instabilities~\cite{Hamdi2016,Ati-PhysRevB.105.014107}, in line with its complex sequence of structural phase transitions with temperature. 
The largest instability is a polar mode ($\Gamma^-_4$) at the Brillouin-zone center that could make WO$_3$ ferroelectric. 
However, the ground state and room temperature phases are both non-polar and show instead $P2_1/c$ and $P2_1/n$ symmetries~\cite{Hamdi2016}. 
These phases can be described as small distortions $\vert \Delta\rangle$ with respect to the cubic phase, that arises from the condensation of distinct unstable modes $\vert \eta_i^c\rangle$ of this cubic phase (i.e. $\vert \Delta \rangle = \sum_i Q_i \vert \eta_i^c\rangle$). 
In the ground-state $P2_1/c$ phase~\cite{Hamdi2016}, the distortion $\vert \Delta_{GS}\rangle$ is dominated by the condensation of (i) unstable antipolar motions of W against O atoms along the $z$ axis, $\vert \eta_M^c\rangle$ ($M^-_3$ mode lowering the symmetry from $Pm\bar{3}m$ to $P4nmm$), and (ii) additional unstable antiphase rotations of the O octahedra, $\vert \eta_R^c\rangle$ ($R^+_4$ mode), that further produce together the appearance of (iii) secondary antipolar motions in the $xy$ plane, $\vert \eta_X^c\rangle$ ($X_5^-$ mode), through improper coupling~\cite{Hamdi2016}. 
In the room temperature $P2_1/n$ phase~\cite{Hamdi2016}, the distortion $\vert\Delta_{RT}\rangle$ involves similar modes, but differently oriented, and also includes additional in-phase oxygen rotations ($M_3^+$ mode), which importantly reduce the $X_5^-$ in-plane antipolar motions along $x$ direction. 
Our approach nicely reproduces the geometry of both phases (see Ref.~\cite{Ati-PhysRevB.105.014107} using the same method) and describes the $P2_1/c$ ground state as an insulator with a bandgap of 3.4 eV, 
in close agreement with the experimental value ($E_g \approx 3.25-3.4$ eV~\cite{band,SIMCHI2014609}).   

\begin{figure*}

\hspace*{-1.5cm}
\centering
\includegraphics[width=1.1 \textwidth]{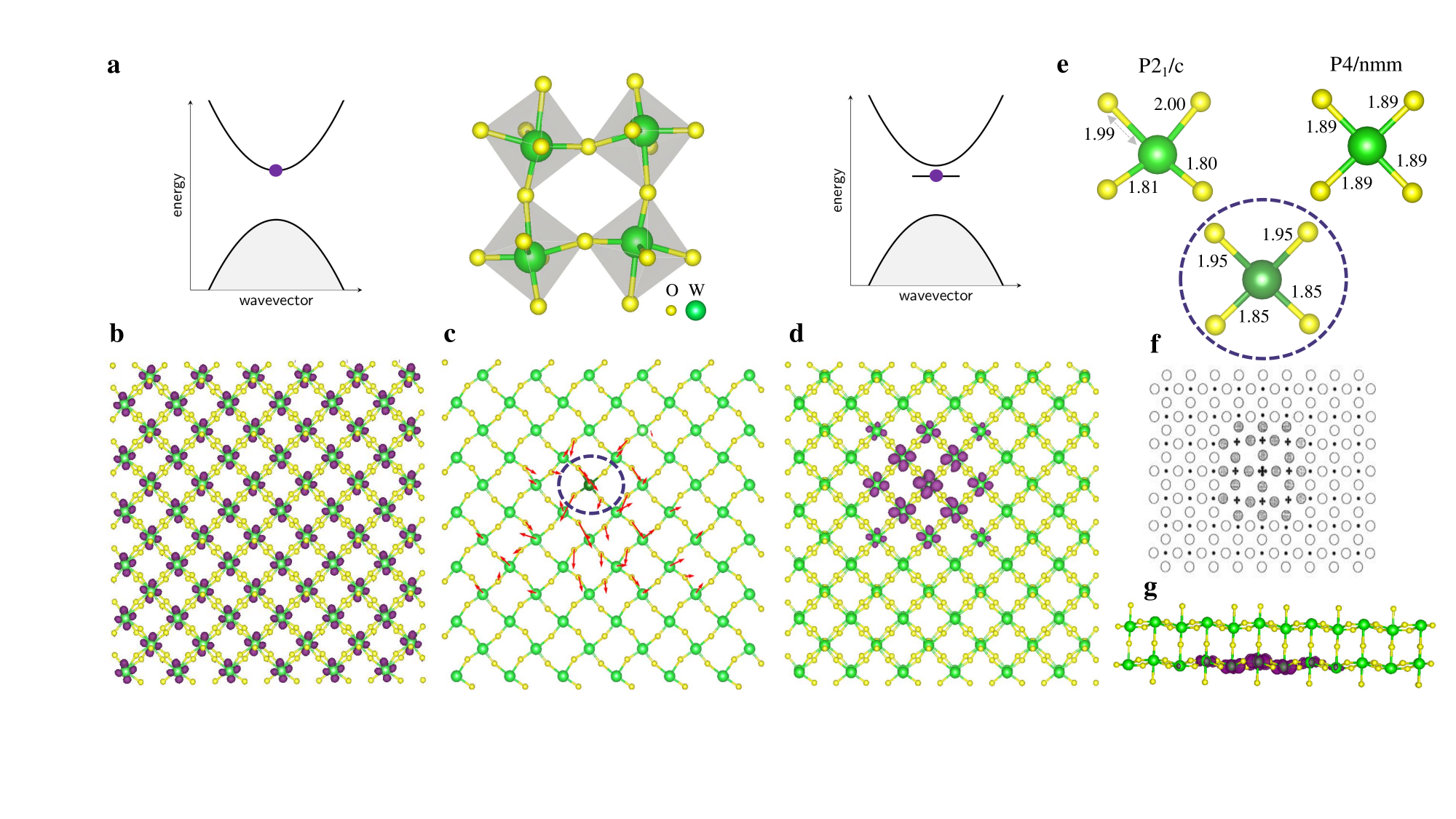}
\caption{\label{fig:1} Polaron in the $P2_1/c$ phase of WO$_3$ (W atoms in green and O atoms in yellow). (a) Sketch of the atomic structure (center) and of the electronic bands, when adding an extra electron, in the free carrier (left) and polaron (right) cases. (b) Computed charge density map (in purple) of a fully delocalized extra electron in the $xy$ plane of the $P2_1/c$ phase. (c) Computed atomic distortion (red arrows) associated to the formation of a single polaron in the $P2_1/c$ phase. (d) Computed charge density map (in purple) of a single polaron in the $xy$ plane of the $P2_1/c$ phase. (e) Comparison of the atomic environment of the polaron central W atom in Fig 1(c) with that in the pristine $P2_1/c$ and $P4/nmm$ phases. (f) Experimental charge density map of a single polaron in the $P2_1/c$ phase as reported by Salje~\cite{Salje-Main-Bipolaron}. (g) Side view of the computed charge density (in purple) of a single polaron in the $P2_1/c$ phase, highlighting its 2-dimensional character.}
\end{figure*}

Adding an extra electron to the system, it should {\it a priori} occupy a state at the bottom of the conduction band, delocalized over the whole crystal (the left panel of Fig.~\ref{fig:1}a, and Fig.~\ref{fig:1}b). 
However, and consistent with previous calculations~\cite{Eric-PhysRevResearch.2.012052}, the proper coupling of this original state with atomic relaxations leads to the formation of a polaron level below the conduction band edge, with the charge located mainly on a central W atom and spreading partly over the first and second W neighbors (the right panel of Fig.~\ref{fig:1}a, and Fig.~\ref{fig:1}d). 
In line with experimental expectations by Salje~\cite{Salje-Main-Bipolaron} (Fig.~\ref{fig:1}f), the polaron charge is confined in a single $xy$ plane (Fig.~\ref{fig:1}g) showing a strongly anisotropic and characteristic 2D disk-like shape  (radius $R_p \approx 7.5-8.5 \, $\AA).
This unusual shape originates from the fact that, in the $P2_1/c$ phase, the lowest conduction level giving rise to the polaron state is made of strongly directional $d_{xy}$ orbitals. Distinctly from the original report~\cite{Eric-PhysRevResearch.2.012052}, our more converged calculations in a bigger supercell locates the polaron state 520 meV below the delocalized state (see also Appendix.~B), supporting the spontaneous formation of polarons at low temperature in the $P2_1/c$ phase of WO$_3$.

\begin{figure*}[h]
\includegraphics[width=1 \textwidth]{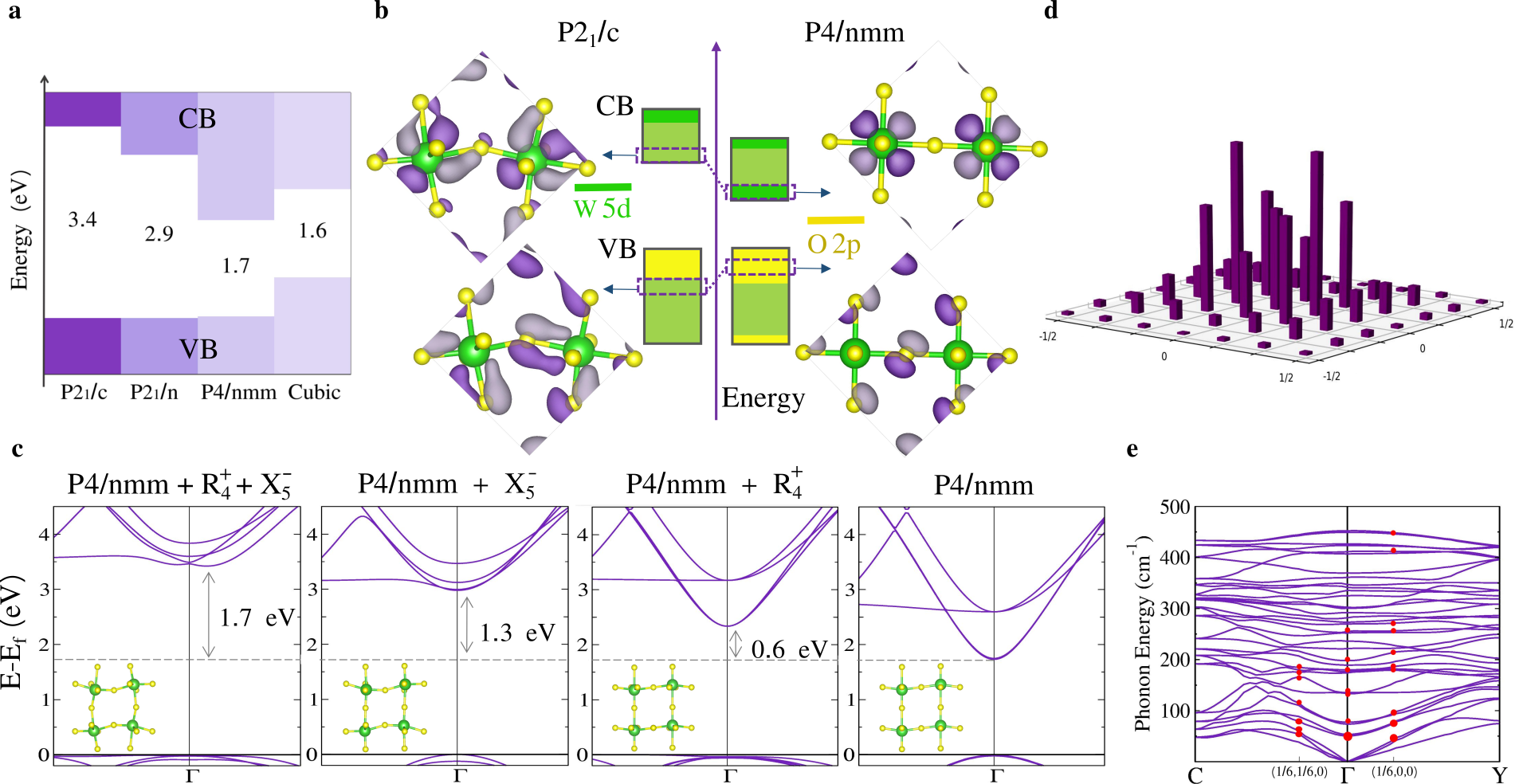}
\caption{\label{fig:2} Electronic and phonon properties of WO$_3$ (W atoms in green and O atoms in yellow). (a) Comparison of the bandgap and position of the valence band (VB) and conduction band (CB) edges in distinct phases of WO$_3$. (b) Simple sketch of the valence and conduction bands in the $P2_1/c$ (left) and $P4/nmm$ (right) phases of WO$_3$, with yellow and green area highlighting respectively the positions of O $2p$ and W $5d$ states. The change of hybridization between O $2p$ and W 5$d_{xy}$ orbitals, yielding the gap opening under atomic distortions from $P4/nmm$ to $P2_1/c$, are illustrated from the evolution of wavefunctions valence and conduction states at $\Gamma$. (c) Evolution of the electronic band structure of the $P4/nmm$ phase of WO$_3$ when condensing the $X_5^-$ in-plane antipolar motion, and $R_4^+$ oxygen rotation motions, bringing the system toward the $P2_1/c$ phase. (d) Contributions to the polaron atomic distortion of phonon modes at distinct ($q_x$, $q_y$, 0) points of first Brillouin zone of the $P2_1/c$ phase (using a $6 \times 6 \times 1$ supercell). (e) Contributions to the polaron atomic distortion of distinct phonon modes from $\Gamma$ and $\Gamma-C$ and $\Gamma-Y$ lines of first Brillouin zone of the $P2_1/c$ phase. Amplitudes of respective contributions are proportional to the red circles' radii.}
\end{figure*}

While the polaron charge is rather isotropically distributed in the $xy$ plane (flat disk-like shape, Fig.~\ref{fig:1}d), the related atomic distortion $\vert \Delta_{pol} \rangle$, illustrated in Fig.~\ref{fig:1}c, is intriguingly asymmetric. 
In order to clarify the nature of $\vert \Delta_{pol} \rangle$, we can project it on the phonon modes $\vert \eta_i^m \rangle$ of the monoclinic $P2_1/c$ phase. 
The dominant contributions are highlighted in Figs.~\ref{fig:2}d,e and are distributed all over the Brillouin zone. 
Amazingly, focusing first on the Brillouin-zone center ($\Gamma$ point), we do not see any significant involvement of polar, Jahn-Teller, or acoustic modes as expected from usual Fr\"olich, Jahn-Teller or Holstein models (see also Appendix.~C). 
Instead, the major contribution at $\Gamma$ comes intriguingly from non-polar Raman modes $\vert \eta_{Ra}^m \rangle$. 
They are these modes that are mainly responsible for the atomic distortion pattern in the polaron core region, but which alone would reproduce this Raman distortion homogeneously to the whole crystal (see Appendix.~C). 
Then, additional non-$\Gamma$ phonon modes are also activated: those yield a global distortion that cancels out the Raman-type distortion outside the core region (see Appendix.~C) and are so responsible for the local character of $\vert \Delta_{pol} \rangle$. 
As a whole, $\vert \Delta_{pol} \rangle$ can therefore be seen as a zone-center Raman-type distortion produced by $\vert \eta_{Ra}^m \rangle$, but localized over only a few unit cells around the polaron core.

At a more atomistic level, the W environment is asymmetric in the pristine $P2_1/c$ phase and, in practice, the polaron distortion acts in order to make this environment more symmetric (Fig.~\ref{fig:1}e). 
Projection of $\vert \eta_{Ra}^m \rangle$ on the phonons of the cubic phase identifies it as a linear combination of $\vert \eta_{R}^c \rangle$ and $\vert \eta_{X}^c \rangle$ inherent to the $P2_1/c$ phase distortion $\vert \Delta_{GS}\rangle$  but oriented in order to cancel it (see Appendix.~C). 
As such, the polaron formation here should not be seen as the activation of an additional virtual phonon cloud attempting to screen the extra charge as in traditional models but, instead, as the local undoing of the $\mid \eta_{R}^c \rangle$ and $\mid \eta_{X}^c \rangle$ distortive motions, originally present in the pristine $P2_1/c$ phase, in order to drive the system closer to the more symmetric $P4/nmm$ phase ($Q_R=Q_X=0$). As such, and in analogy with {anti-Jahn-Teller} polarons~\cite{Anti-Jahn-Teller}, we coin here the name of {\it anti-distortive} polaron.

Looking at the various metastable phases of WO$_3$ \cite{Hamdi2016}, it appears in Fig.~\ref{fig:2}a that the electronic bandgap is much larger in the  $P2_1/c$ (3.4 eV) and $P2_1/n$ (2.9 eV) phases than in the $P4/nmm$ phase (1.7 eV), which is in line with previous theoretical and experimental reports~\cite{Hamdi2016,Ping-14} and can be rationalized as follows (see also Appendix.~D).
The bandgap in WO$_3$ is typically between valence and conduction states respectively of dominant O $2p$ and W $5d$ $t_{2g}$ character~\cite{Wang2011}. 
In the reference cubic phase, all three $t_{2g}$ states are degenerated at the bottom of the conduction band.
In the $P4/nmm$ phase in which all in-plane W-O bonds are equivalent, the condensation of $M_3^-$ antipolar motions along $z$ has pushed up $d_{xz}$ and $d_{yz}$ states so that the lowest conduction states are of $d_{xy}$ character. 
Then, the appearance of $X_5^-$ antipolar in-plane motion, and to a lower extent $R_4^+$ oxygen rotation motions, create alternating short and long W-O bonds in-plane. 
This amplifies hybridizations between the occupied O $2p$ and empty W $5d_{xy}$ states and significantly opens the bandgap by lowering, in the valence band, the occupied state localizing the charge in the short W-O bond and by moving up, in the conduction band, the empty state localizing the charge in the long W-O bond  (Figs.~\ref{fig:2}b,c). 
Our conjecture is therefore that, in order to form the polaron, the system suppresses locally the $X_5^-$ and $R_4^+$ atomic distortion inherent to the pristine $P2_1/c$ phase in order to lower the conduction state there and trap the extra charge at a lower energy. This mechanism is reminiscent of the phase transformations previously reported in WO$_3$ over heavy doping \cite{vandewalle-17}.

As a first proof of concept, we built a supercell of the $P2_1/c$ phase, in which we artificially imposed $Q_R=Q_X=0$ around the central unit cell (see Appendix.~E) to bring the system back there to $P4/nmm$. 
Adding an extra electron to that frozen structure, it naturally localizes in the central region, showing features similar to the relaxed polaron (see Appendix.~E). 
This confirms the possibility to achieve charge localization from the tuning of Raman mode distortions but it does not demonstrate that such a mechanism will spontaneously appear.

In order to go one step further, we developed a simple quantum-dot model~\cite{Kouwenhoven-01} (see Appendix.~F). 
Adding an extra electron to the system, it should {\it a priori} occupy the lowest conduction energy level and delocalize over the whole crystal (Fig.~\ref{fig:1}b), which defines our reference energy state. 
Starting from there, forming a polaron requires activating the atomic distortion $\vert \Delta_{pol}\rangle$ in the dynamically stable $P2_1/c$ phase. 
This costs an energy $E_{latt}(x,R)$ which scales both with the amplitude $x$ of the Raman-type atomic distortion $\vert \eta_{Ra}^m \rangle$ associated to the polaron (with $x$ normalized to 1 in the polaron structure relaxed in DFT) and the radius $R$ of the region within which this distortion appears. 
Concomitantly, condensing a Raman distortion $\vert \eta_{Ra}^m \rangle$ of amplitude $x$ in the $P2_1/c$ phase lowers the lowest conduction state containing the extra electron, producing then a lowering of electronic energy $E_{elec}(x)$. 
The latter is an electronic band term, expected when condensing $\vert \eta_{Ra}^m \rangle$ homogeneously over the whole crystal. 
Restricting the distortion over a spacial region of radius $R$ will further localize the extra electron in this region and so produce an extra energy cost $E_{conf}(x,R)$ due to the confinement of the electron. 
In this context, the polaron formation energy can be written as 
\begin{equation}
 E_{f}^{pol} = \min_{(x,R)} \{ E_{latt}(x,R) +E_{elec}(x) +E_{conf}(x,R) \}
\end{equation}
and the polaron will form spontaneously as soon as the lowering of $E_{elec}$ produced by the appearance of the Raman distortion compensates the energy costs produced by $E_{latt}$ and $E_{conf}$. 

The explicit $(x,R)$ dependence and parameters of all three energy terms can be independently estimated from DFT calculations (see Appendix.~F). 
Then, determining $E_f^{pol}$ from the minimization in Eq. (1) will not only tell us if the polaron will naturally appear (i.e. $E_{f}^{pol}<0$) but also what is the spatial extension $R$ and distortion amplitude $x$ at the minimum. 
 As discussed in Appendix.~F, the model calculations reproduce the formation of a polaron over a relatively broad range of effective masses entering into $E_{conf}(x,R)$.
Moreover, a formation energy ($E_f^{pol} \approx -100$ meV), polaron radius ($R_p \approx 8 \;$\AA) and distortion amplitude ($x \approx 1$) comparable to DFT data are obtained consistently for realistic values of the parameters.

The ability of this simple quantum dot model to reproduce, with realistic parameters, the results achieved from the fully self-consistent DFT atomic relaxation, provides strong support to our initial conjecture that polaron formation in the $P2_1/c$ phase of WO$_3$ arises from local lowering of the bandgap achieved from the partial undoing of the atomic distortions inherent to the pristine phase. 
According to this picture, the distinct phases of WO$_3$ should not be similarly prone to form polarons.
Consistently with that, in the room-temperature $P2_1/n$ phase, which shows smaller $Q_X$ along $x$ and a lower bandgap, polarons become much more delocalized in our calculations (see Appendix.~G), as also confirmed experimentally~\cite{Salje-Main-Bipolaron}. 
Then, in the $P4/nmm$ phase ($Q_R=Q_X=0$), we were unable to stabilize any similar polaronic state.

\noindent
\section{Discussion}

It is worth contrasting our anti-distortive polaron with traditional conceptual models. 
In WO$_3$, the strong polar character of the compound (giant Born effective charges $Z^*$~\cite{Detraux97}, soft polar mode in the cubic phase) was naturally pointing in the direction of an efficient dielectric screening of the extra charge through the activation of longitudinal optical polar modes in line with Landau-Pekar-Fr\"olich models or of a ferroelectric polaron. 
However, our DFT calculations show that no significant polar motion is involved in the polaron distortion. 
Alternatively, WO$_3$ might also be prone to form Jahn-Teller polarons due to the Jahn-Teller active character of the W ion. 
However, the polaron distortion in the $P2_1/c$ phase does not overlap with Jahn-Teller motions either. 
The present anti-distortive polaron of WO$_3$ shows closer similarities with small Holstein polarons in the sense that charge trapping relates to the formation of a quantum dot. 
However, while the conventional Holstein polaron is typically linked to the activation of a local elastic response induced by the extra charge, here the mechanism involves changes of hybridizations driven by the undoing of soft atomic motions inherent to the displacive nature of the compound. The unusual {\it medium} size of the polaron in WO$_3$ supports that both mechanisms are not necessarily strictly equivalent. 

The case of WO$_3$ could be a bit specific since the absence of the A cation gives rise to large distortions and the 5$d$ character of W makes the $d$ states particularly delocalized, but the mechanism of anti-distortive polaron is {a priori} totally generic and should more broadly apply to various displacive compounds like some related ABO$_3$ perovskites. The latter are typically mixed ionic-covalent compounds and their giant $Z^*$ testify of the strong sensitivity of orbital hybridization to atomic displacements~\cite{Ghosez98}. Depending on their Goldschmidt tolerance factor~\cite{Goldschmidt}, ABO$_3$ compounds are prone to develop different kinds of structural distortions~\cite{Ghosez22}, which are known to tune their bandgap, and could get similarly involved in polaron formation. 

As a guiding rule, inspection of the electronic bandgap and its evolution from phase to phase might be a good indicator to track the eventual propensity of alternative compounds to develop analogous anti-distortive polarons. A quick search in materials databases for compounds showing large increase of bandgap from a high-symmetry to a lower-symmetry phase, provides a long list of perovskite and even non-perovskite potential candidates, including WO$_3$ (see Appendix.~H). All of them might not be relevant since other aspects need to be considered, such as the nature of the bandgap. 
Nevertheless, considering for instance $P2_1/c$ MoO$_3$ (isostructural to WO$_3$) or  $Pnma$ YAlO$_3$ and $P2/c$ YTaO$_4$ as representative members of the perovskite $R$(Al/Ga)O$_3$ and fergusonite $R$(Ta/Nb)O$_4$ families of compounds (with $R$ = Y or a rare-earth) and adding an extra electron, we can stabilize in those compounds an anti-distortive electronic polaron (see Appendix.~H) arising from the partial undoing of distortive motions, exactly as in WO$_3$. This illustrates concretely the generality of the reported mechanism that might also be relevant to many other families of displacive compounds.

While anti-distortive polarons emerge from the undoing of distortive motions that had opened the electronic bandgap in the pristine phase, we could imagine alternatively distortive polarons activating instead a distortion lowering the conduction states. 
This points out some interesting connections with (anti)Jahn-Teller polarons~\cite{muller,Lenjer02,Anti-Jahn-Teller}. 
But while in Jahn-Teller polarons the tuning of the gap arises from on-site splitting of degenerated states from the Jahn-Teller effect, in anti-distortive polarons, it is linked to inter-site covalency effects and the tuning of bonding-antibonding states. 
It is worth noticing that the present discussion also links to the metal-insulator transition in rare-earth nickelates in which the activation of distortive breathing oxygen motions lower energy states to localize the charge at some sites and is sometimes interpreted as a polaron condensation~\cite{Shamblin18}. 

In conclusion, the present concept of anti-distortive polaron is not only of academic interest. Appearing relevant to various compounds, it naturally connects to various fields including electrochromism \cite{DEB2008245}, high-T$_c$ superconductivity \cite{muller,Shengelaya20}, colossal magneto-resistance ~\cite{1997Natur.386..256T,PhysRevB.54.5405,Shengelaya20}, photovoltaics \cite{Photovoltaics}, photocatalysis \cite{Photocatalysis} or ion batteries \cite{Batteries}. Moreover, the distinct behaviors highlighted in the $P2_1/c$, $P2_1/n$ and $P4/nmm$ phases of WO$_3$ (see Appendix.~G) demonstrate explicitly the strong sensitivity of the polaronic state with the amplitude of distortive motions. In WO$_3$ \cite{Mazzola-23}, related perovskites \cite{Ghosez22} and other displacive compounds, phase stability and distortive motions can nowadays be efficiently monitored by strain~\cite{Haeni-04} and interface~\cite{Kim-16,Liao18} engineerings, electric fields~\cite{Varignon16} or optical pulses~\cite{Nova19}. 
So, linking explicitly here polarons in some compounds to tunable distortive motions, we open {\it de facto} concrete perspectives to achieve in those compounds rational static or dynamical control of the polaronic state and of its related properties by external fields. 

\section*{Appendix}

This article has an accompanying appendix, which is composed of eight sections to which we refer in the main text for supporting different claims.

\section*{Acknowledgements}

P.G. thanks Elaheh Ghorbani for useful discussion. This work was supported by F.R.S.-FNRS Belgium (grant PROMOSPAN). The authors acknowledge access to the CECI supercomputer facilities funded by the F.R.S-FNRS (Grant No. 2.5020.1) and to the Tier-1 supercomputer of the Fédération Wallonie-Bruxelles funded by the Walloon Region (Grant No. 1117545).

\section*{Authors contributions}
P.G. conceived the study and supervised the work with B.P. . H.H. performed the calculations and analyzed the results. X.H. and H.H. made the search and characterization of alternative compounds. E.B. provided support for the first-principles calculations and B.P. for the development of the quantum dot model. P.G. and H.H. wrote the manuscript. All authors discussed the results during the project and commented the manuscript.

\section*{Declarations}

The authors declare no competing interest.

 \section*{Data Availability}
 The datasets generated and analyzed in this study are available from the corresponding author upon request.

 \section*{Code Availability}
 The data presented in this study were generated using broadly accessible first-principles packages as described in the Methods section in Appendix.

\clearpage
\bibliography{Polaron.bib}

\end{document}